\documentclass[apjl]{emulateapj}
\usepackage{apjfonts}

\received{July 24, 2008}
\accepted{August 20, 2008}

\newcommand\teff{\ensuremath{T_{\rm eff}}}

\slugcomment{Accepted for publication in the Astrophysical Journal Letters} 

\shorttitle{Metal Inhomogeneity and Settling in Pulsating WDs}

\shortauthors{Montgomery, Thompson, \& von Hippel}

\begin{document}
\title{Constraining the Surface Inhomogeneity and Settling Times of
  Metals on Accreting White Dwarfs}

\author{M.\ H.\ Montgomery\altaffilmark{1,3}, S. E. Thompson\altaffilmark{2,3}, 
  T. von Hippel\altaffilmark{1,4}}

\altaffiltext{1}{Department of Astronomy, University of Texas at
  Austin, Austin, TX, USA; mikemon@astro.as.utexas.edu}
\altaffiltext{2}{Department of Physics and Astronomy, University of 
  Delaware, Newark, DE}
\altaffiltext{3}{Delaware Asteroseismic Research Center, Mt.\ Cuba
  Observatory, Greenville, DE} 
\altaffiltext{4}{Department of Physics, Siena College, Loudonville, NY}

\begin{abstract}
  Due to the short settling times of metals in DA white dwarf
  atmospheres, any white dwarfs with photospheric metals must be
  actively accreting. It is therefore natural to expect that the
  metals may not be deposited uniformly on the surface of the star. We
  present calculations showing how the temperature variations
  associated with white dwarf pulsations lead to an observable
  diagnostic of the surface metal distribution, and we show what
  constraints current data sets are able to provide. We also
  investigate the effect that time-variable accretion has on the metal
  abundances of different species, and we show how this can lead to
  constraints on the gravitational settling times.
\end{abstract}

\section{Astrophysical Context}

There are two main classes of white dwarf stars: those with
hydrogen-rich atmospheres (spectral type DA) and those with
helium-rich atmospheres (non-DA spectral types). The reason for this
is that the high surface gravities of white dwarfs lead to efficient
gravitational settling, with the lightest elements rising to the
surface. In addition, some DA white dwarfs have spectra showing metal
lines of elements such as Ca and Mg \citep{Zuckerman03}, and these
stars are referred to as DAZs; about 20\% of all DAs fall into this
category. Recently, \citet{Dufour07} have announced a new class of
white dwarf with carbon-dominated atmospheres, the ``hot DQ'' stars,
several examples of which have been found in the Sloan Digital Sky
Survey \citep{Liebert03}.



The presence of metals in the DAZs is intriguing since the settling
time scale for the metals may be many orders of magnitude shorter than
the evolutionary age of these objects.  Indeed, for DAZs with $\teff
\sim$~12,000~K, the settling time scale can be on the order of days or
weeks, meaning that these objects are experiencing ongoing accretion
\citep{Koester06}.  This ongoing accretion is consistent with the fact
that nearly a dozen of these objects have detected dust disks
\citep[``debris disks,''
see][]{Tokunaga90,Becklin05,Kilic05,Reach05,Kilic06,vonHippel07a,Jura07a,Jura07b,Farihi08}
and these disks are assumed to be the sources of the metal lines seen
in these white dwarf atmospheres.  The best studied object of this DAZ
class with an observed disk, G29-38, is also a multi-periodic variable
white dwarf (DAV), pulsating in non-radial g-modes with periods of a
few hundred to one thousand seconds.

The technique of asteroseismology uses the observed pulsation modes of
a star to infer and constrain the interior structure of the star, thus
obtaining information on the star's structure as a function of radius
\citep{Bradley94a,Kawaler94,Metcalfe02,Montgomery03}.  In contrast to
this, our approach in this paper is to use the different
\emph{angular} dependence of the pulsations to constrain the accretion
process. Since the accretion is most likely occurring through a disk,
it is natural to suppose that the metals may not be uniformly
distributed across the star's surface. We present calculations showing
how we can place constraints on the non-uniformity of the accretion
process.

\section{Gravitational Settling and Horizontal Diffusion}

For all our calculations, we use the DAZ G29-38 as our template,
since, given its brightness, it has the greatest potential for
successful measurement of a surface inhomogeneity. In addition, we have
archival data on this star appropriate to this application.  In this
section, we will therefore attempt to quantify the importance of
gravitational settling and horizontal diffusion assuming a model with
parameters similar to those of G29-38.

\begin{figure}[b]
  \centering{
    \includegraphics[height=1.0\columnwidth,angle=-90]{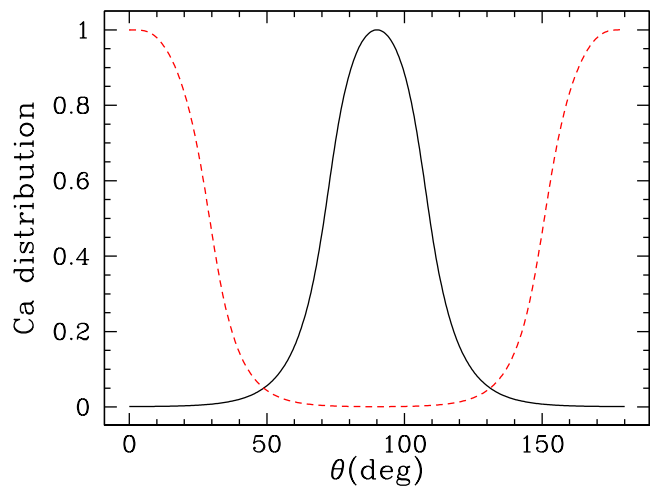}
  }
  \caption{Possible surface metal distributions for accretion
    centered on the poles (dashed curve) or the equator (solid curve)
    as a function of the polar angle (co-latitude).}
  \label{zdist}
\end{figure}

\begin{figure*}[t]
\includegraphics[angle=-90,width=0.5\textwidth]{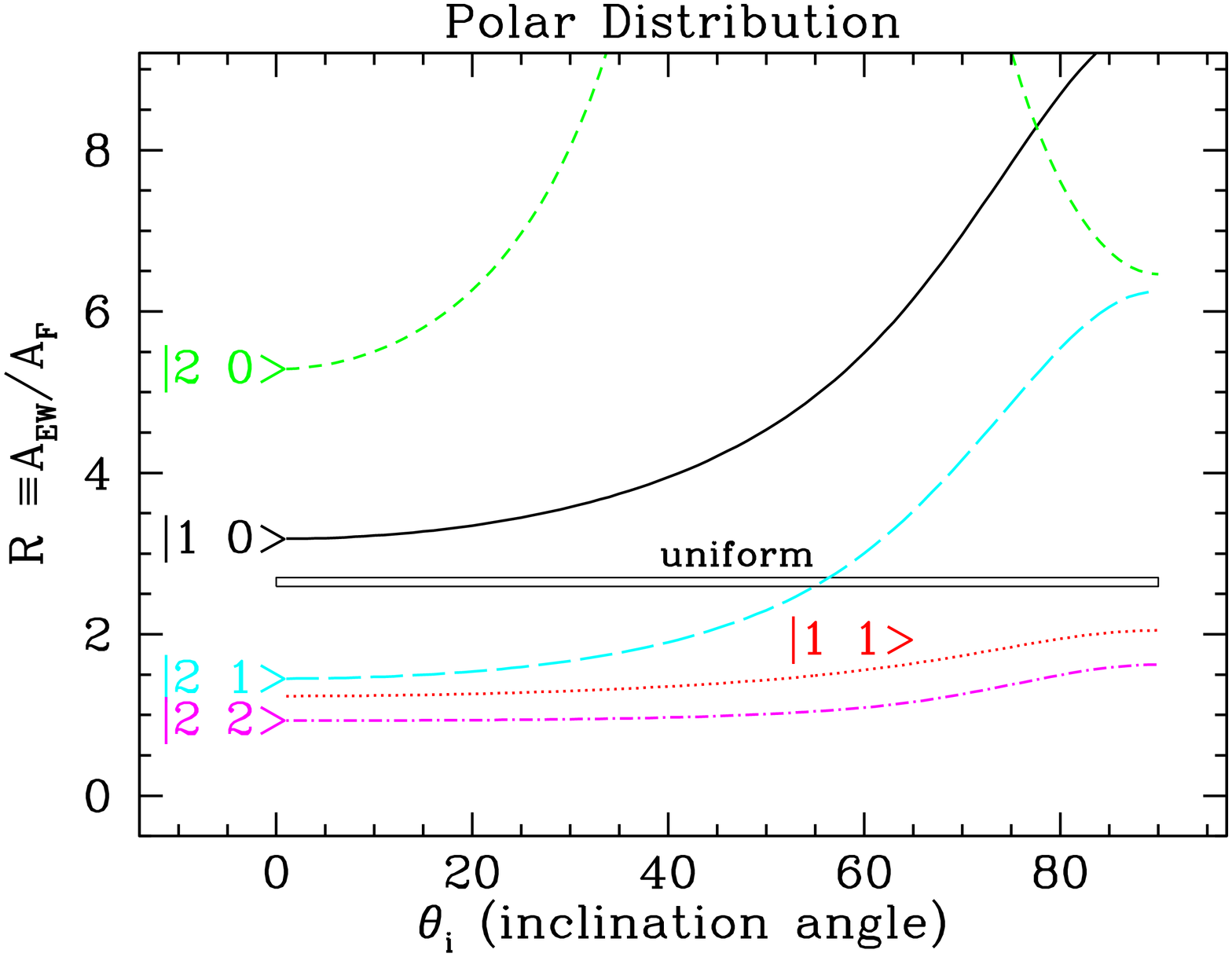}
\includegraphics[angle=-90,width=0.5\textwidth]{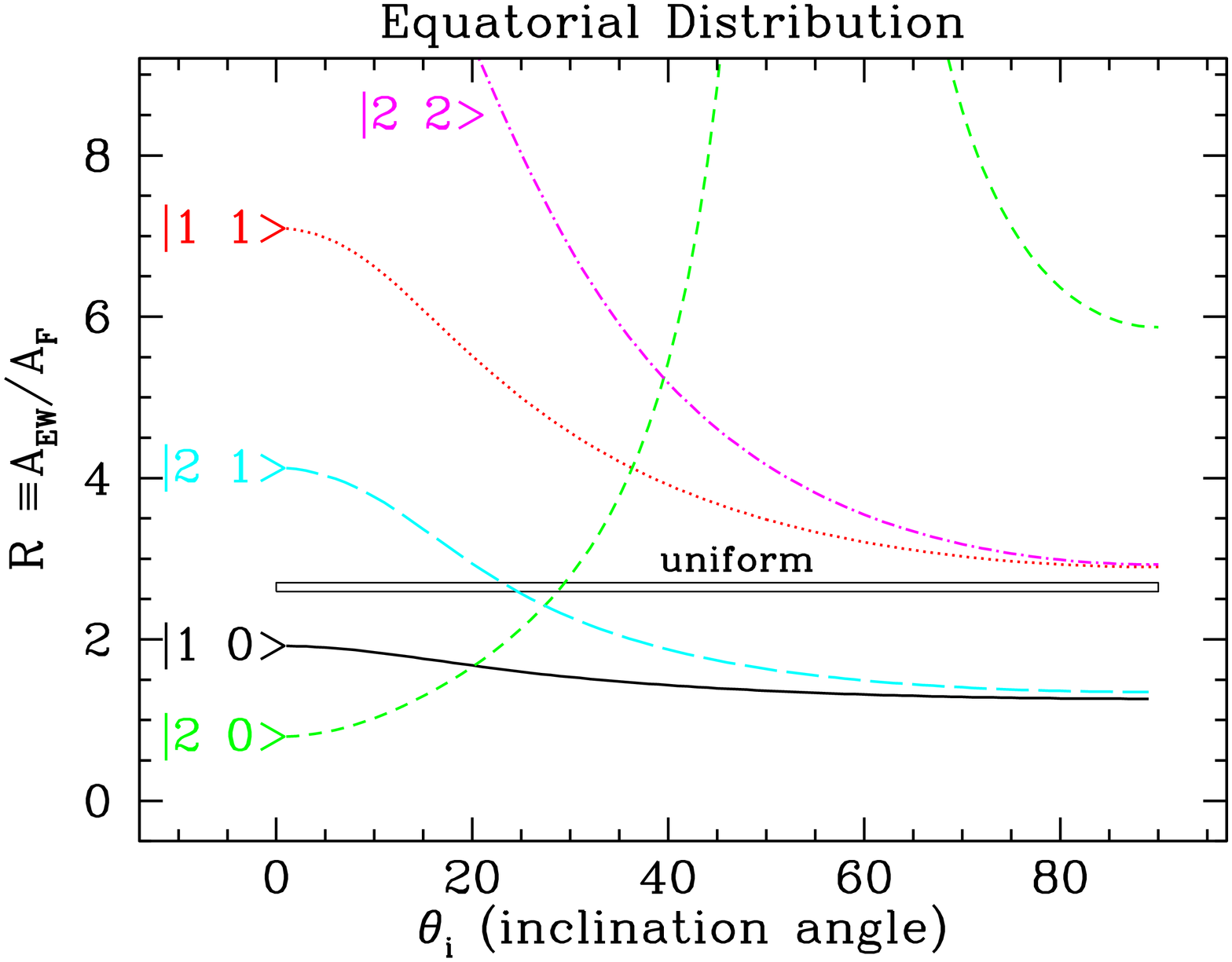}
\caption{ The diagnostic $R$ as a function of inclination angle
  $\theta_{\rm i}$. The left panel shows the results for the polar
  distribution of metals shown in Figure~\ref{zdist} and the right panel
  shows the results for the equatorial distribution. The different
  curves are labeled by the $| \,\ell\, m \,\rangle$ values of the
  relevant pulsation modes.
  \label{rdiag}}
\end{figure*}

\citet{Bergeron04} find $\teff =$~11,820~K and $\log g = 8.14$ for
this star, while \citet{Koester97} find $\teff =$~11,600~K and $\log g
= 8.05$. Interpolating in the tables of Koester \& Wilken for Ca
yields a settling time of $\sim 13$ days for the first set of
parameters\footnote{Due to a mis-reading of Table~2 in
  \citet{Koester06}, \citet{vonHippel07} calculated a settling time of
  $\sim 7$ days.}  and a settling time of $\sim 23$ days for the
second. We note that while unseen helium in the atmosphere could
lengthen these settling times considerably
\citep[e.g.,][]{Garcia-Berro07}, its presence is inconsistent with the
pulsation results for this star: such an amount would imply a
\emph{much} deeper surface convection zone, in conflict with that
found by \citet{Montgomery05a}.

Since these stars should have surface convection zones, the dominant
form of horizontal transport of Ca will be due to the turbulent
viscosity. We can estimate the size of this diffusion coefficient as
$D \sim v_C l_h$, where $v_C$ is a typical convective velocity and
$l_h$ is the assumed ``mixing length'' for convection. From our white
dwarf evolution code \citep[e.g., see ][]{Montgomery99} we find for
both sets of stellar parameters that $D \approx 1.5\cdot 10^{10} {\rm
  cm}^2/{\rm sec}$.

For these simple exploratory calculations we assume azimuthal symmetry
for both the accretion and the surface metal distribution, i.e., $Z =
Z(\theta,t)$ and $S = S(\theta,t)$, where $Z(\theta,t)$ is the metal
abundance, $S(\theta,t)$ is the source function of metals accreting
onto the white dwarf, $\theta$ is the ``co-latitude'' of a point on
the star's surface, and $t$ is time.

Since convection will uniformly mix material vertically, we can treat
the convective region as a single ``zone'' and write an equation for
the time evolution of $Z$ as a function of $\theta$ and $t$:
\begin{equation}
  \frac{\partial Z(\theta,t)}{\partial t} = - \gamma \, Z(\theta,t)
  + D \, \nabla^2_{\rm h} Z(\theta,t) + S(\theta,t),
  \label{zeqn}
\end{equation}
where $\gamma \equiv 1/\mbox{(settling time)}$ is the settling rate,
$\nabla^2_{\rm h}$ is the horizontal part of the Laplacian operator,
and the other variables are as defined above. An estimate of the
relative importance of sinking to spreading is $\eta \equiv \gamma \,
R^2_\star/D$, where $R_\star$ is the radius of the white dwarf: $\eta
\sim 50$ using the \citet{Bergeron04} values whereas $\eta \sim 30$
for those of \citet{Koester97}. When $\eta \gg 1$ the metals will sink
before they have a chance to diffuse horizontally, while for $\eta \ll
1$ the metals will have a chance to mix thoroughly horizontally before
sinking, producing a nearly uniform surface distribution.

In Figure~\ref{zdist} we show the metal distributions which arise from
solutions of equation~\ref{zeqn}. The dashed curve is the equilibrium
distribution which results from constant accretion centered at the
poles and the solid curve is that which results from constant
accretion centered on the equator.

\section{The Diagnostic}

The flux variations observed in pulsating white dwarfs are due almost
entirely to temperature changes on the surface of the stars
\citep{Robinson82}. These same temperature changes will also affect
the equivalent widths (EWs) of any spectral lines, and in particular
the EWs of metal lines. These metals may not be uniformly distributed
across the star's surface and since the temperature variations are
also non-uniform, we hope to be able to constrain the surface metal
distribution.

The relevant diagnostic we have developed, denoted by $R$, is
the ratio of the fractional EW amplitude to the fractional
flux amplitude:
\begin{equation}
  R \equiv \frac{\delta \langle EW \rangle/\langle EW \rangle}{
    \delta F_X/F_X} = \frac{A_{\rm EW}}{A_{\rm Flux}},
\end{equation}
where $A_{\rm EW}$ is the fractional amplitude of $EW$ variations, and
$A_{\rm Flux}$ is the amplitude of the fractional flux variations
observed in the given passband $X$.

To compute $R$, we assume a particular surface temperature
perturbation of the form $\delta T/T \propto Y_{\ell m}(\theta,\phi)$.
We then use model atmospheres to turn this into EW and flux variations
on the surface of the star, taking into account the fact that the EW
of the Ca lines will be directly proportional to the local abundance
of Ca (e.g., the curves in Figure~\ref{zdist}). For the passband $X$ we
assume a wavelength response appropriate to the Argos CCD with a BG40
filter on the 2.1m telescope at McDonald Observatory \citep{Nather04};
the wavelength range is taken to be 3000~\AA\ to 7000~\AA , with a
peak response at 5400~\AA.  Finally, we integrate the result across
the visible surface of the star. Limb darkening is automatically taken
into account by this procedure.

We note for a uniform distribution of metals that $R=2.71$ for
$\ell=1$ and $R=2.59$ for $\ell=2$. Thus, even by measuring $R$ for a
single mode it may be possible to tell whether metals are uniformly
distributed on the star's surface. To obtain further constraints, we
need $R$ determinations from other modes and/or additional information
such as the inclination angle of the star.

In Figure~\ref{rdiag} we show the $R$ diagnostic as a function of
inclination angle $\theta_{\rm i}$. The different curves are labeled
according to their $\ell$ and $m$ values as $|\,\ell\, m\,\rangle$.
The left-hand plot is for the polar distribution of metals given in
Figure~\ref{zdist} and the right-hand plot is for the equatorial
distribution. The thin horizontal boxed region shows the value of $R$
expected if the metal distribution is uniform; we see that a
non-uniform distribution is very unlikely to produce a value of $R$ in
this range.

\begin{figure}[t]
  \centering{
    \includegraphics[angle=-90,width=1.0\columnwidth]{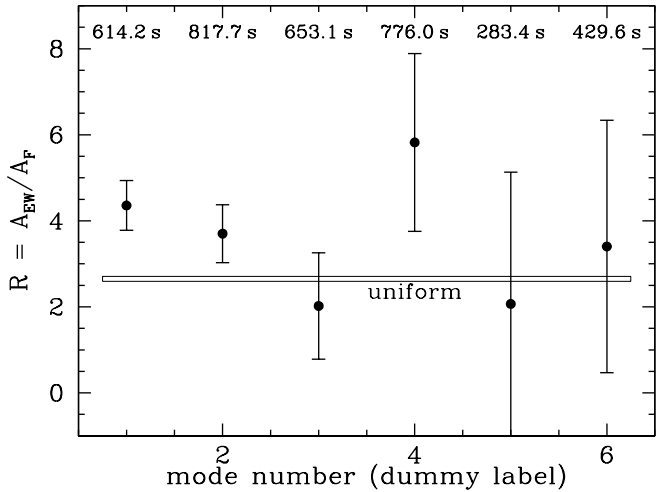}
  }
  \caption{Values of $R$ for modes in the DAV G29-38 as obtained from
    4 hours of Keck data in 1999. The values for the first 6 highest
    amplitude modes are shown, with the period in seconds indicated
    above each point. Only the first two modes have a signal-to-noise
    ratio large enough to be useful.  }
  \label{keck}
\end{figure}

\section{Comparison with Observations}

In 1996 Clemens \& van Kerkwijk obtained over 4 hours of time-resolved
spectroscopy of the DAV G29-38
\citep{Clemens00,vankerkwijk00,Clemens99}. While not originally
intended for this purpose, we can use this as an example data set for
the technique proposed in the previous section. We take the amplitudes
for the EW variations from the analysis of \citet{vonHippel07} and we
use the amplitudes of the broadband (5200--5500 \AA) flux variations
as determined by \citet{vankerkwijk00} \footnote{Even though this
  wavelength range is much narrower than that used in the previous
  section for Figure~\ref{rdiag}, the central wavelengths of the two
  passbands are nearly the same and the results obtained are
  virtually indistinguishable from one another.}. We calculate
$R=A_{EW}/{A_{\rm Flux}}$ for each of the modes, taking into account
the errors on all quantities.

We show the results of this procedure in Figure~\ref{keck}.  Of the 6
modes we identified, only the first two have $R$ values with small
enough error bars to provide any constraint on the Ca distribution.
The 614~s mode provides the most convincing evidence for a non-uniform
Ca distribution, since its $R$ value is a full $3 \sigma$ above the
range produced by a uniform distribution. The 818~s mode may also
provide some evidence, although it is only about $1.3 \sigma$ above
the value expected from a uniform distribution.

\begin{figure}[t]
  \centering{
    \includegraphics[angle=-90,width=1.0\columnwidth]{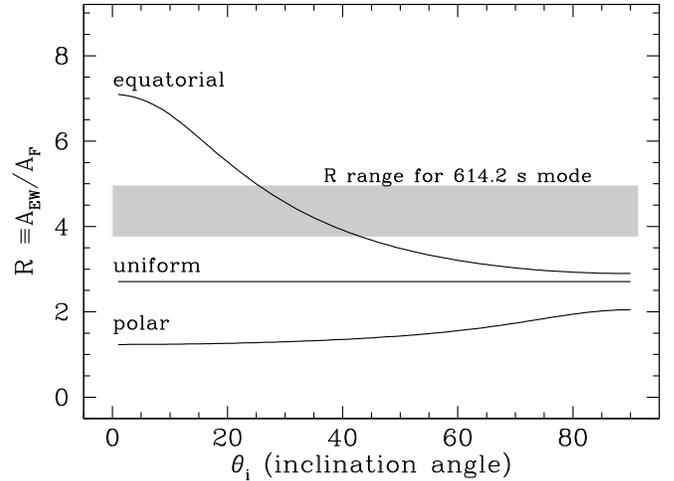}
  }
  \caption{Comparison of the observed $R$ value for the 614~s mode in
    G29-38 (shaded region) to the value expected for a polar, an
    equatorial, and a uniform distribution of Ca, as a function of
    inclination angle $\theta_{\rm i}$. The data are only consistent
    with the equatorial case.  }
    \label{rex}
\end{figure}

With only one statistically significant value of $R$, we cannot hope to
infer anything further about the Ca distribution. Fortunately, we do
have additional information.  \citet{Clemens00} used time-resolved
spectroscopy to determine that this mode has $\ell=1$. In addition,
\citet{Montgomery05a}, by modeling the nonlinear pulse shape of this
mode, was able to further specify that it is an $\ell=1$, $m=1$ mode.
With just this one additional constraint, we are somewhat surprisingly
able to constrain our models of the Ca distribution. Figure~\ref{rex}
shows the $R$ values derived using an equatorial, a polar, and a
uniform distribution, for a variety of inclination angles. The 614~s
measurement is inconsistent with a polar distribution of metals, and
is most consistent with an equatorial distribution.

\section{Time-variable Accretion and Settling}

In the case of G29-38, accretion is thought to be from a debris disk
having roughly asteroidal composition, so that several elements are
being accreted simultaneously. In particular, if we look at Ca and Mg,
these two elements have different settling rates, with Mg settling
about 50\% more slowly than Ca.

Accretion is not expected to be a completely steady process.
\citet{vonHippel07} found the EW of Ca lines in G29-38 to vary on a
time scale of weeks to years, with additional evidence for variations
on shorter time scales (von Hippel, private communication). On the
other hand, \citet{Debes08} found the observed EW variations of G29-38
to be consistent within the measurment errors. For the present we
treat this as an open question and ask what effect time variable
accretion would have on the measured atmospheric abundances of
different chemical species, and how this can be used as a probe of
the settling process in these stars.


\begin{figure}[t]
  \centering{
    \includegraphics[width=1.0\columnwidth]{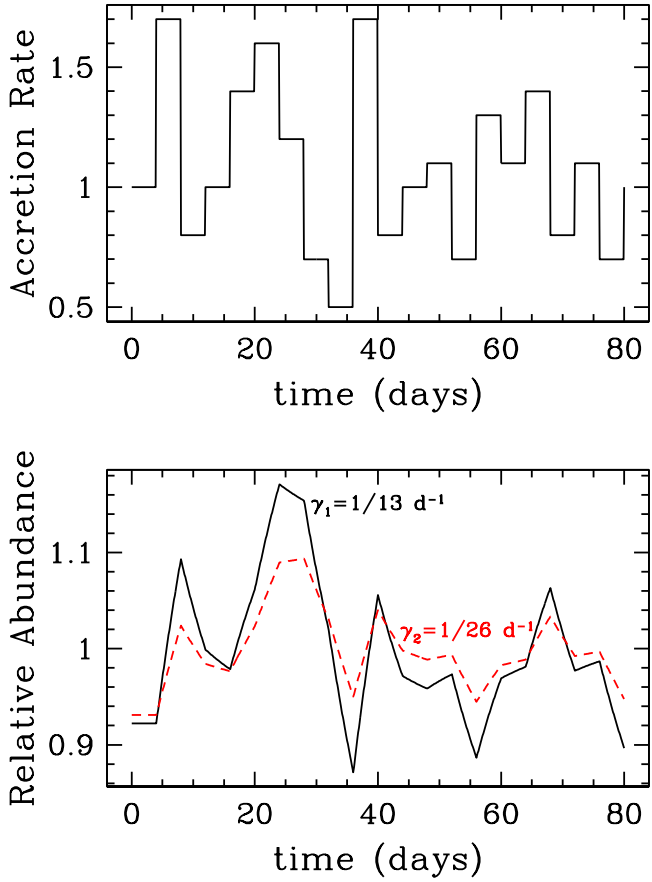}
  }
  \caption{Bottom panel: the fractional relative abundance of element~1 
    ($\gamma_1= 1/13$~d$^{-1}$, solid line) and element~2
    ($\gamma_2=1/26$~d$^{-1}$, dashed line) as given by the solution
    of equation~\ref{zacc}.  Top panel: the time-variable accretion
    rate (arbitrary units) assumed for these calculations.
  \label{varacc}}
\end{figure}

For this problem it is sufficient to consider the spherically averaged
version of equation~\ref{zeqn}:
\begin{equation}
  \frac{d Z_i(t)}{d t} = - \gamma_i \, Z_i(t) + X_i\, S(t),
  \label{zacc}
\end{equation}
where $Z_i$ is the abundance of element $i$, $\gamma_i$ is its settling
rate, and $X_i$ (assumed to be constant) is the relative fraction of
element $i$ in the accreted material.

In Figure~\ref{varacc} we illustrate the solution of
equation~\ref{zacc} for a particular set of parameters and a given
accretion rate. The accretion rate assumed is shown in the top panel
of Figure~\ref{varacc}. For the two elements we have assumed that
$\gamma_1 = 1/13$~d$^{-1}$ and $\gamma_1 = 1/26$~d$^{-1}$. In the
bottom panel we plot the abundance of each element divided by its mean
value, i.e., its ``fractional'' abundance: the solid curve is that of
element 1 and the dashed curve is that of element 2. By using
relative/fractional abundances, we have divided out the effect of the
$X_i$ parameter.

The first feature to note is that the abundances of the two elements
are highly correlated, i.e., when one goes up or down so does the
other. In addition, we see that the amplitude of the excursions of
element~1 are larger than those of element~2.  In fact, taking the
standard deviations of each of the curves we find that that of
element~1 is about 1.8 times that of element~2.  In other words, the
element with the shorter settling time experiences larger fractional
variations, and these variations may be used for a rough estimate of
the ratio of the settling times. Thus, periodic observations of a
white dwarf with more than one metal line may be able to place
constraints on the relative settling times of different chemical
species in white dwarf atmospheres.

\section{Conclusions}

We have shown how the temperature variations due to stellar pulsation
can be used to constrain the metal distribution on the surface of a
white dwarf, and we have shown that data currently in hand for the
star G29-38 suggests that the metal distribution, and therefore the
accretion, may be equatorial. If further studies support an equatorial
distribution for Ca in G29-38, this argues against magnetic accretion
onto spots near the poles.  Non-magnetic equatorial accretion would
further imply that the inner edge of the disk is much thinner than the
white dwarf's radius.  A physically thin disk is consistent with
observations to date \citep[e.g.,][]{Jura07a,vonHippel07a}.

We have also shown how the observed variations in atmospheric
abundance of two chemical species can be used to place constraints on
their settling rates. The element with the faster settling rate will
show the larger fluctuations, and, at least within the context of this
simplified model, the ratio of the magnitude of these excursions from
the mean will be approximately proportional to the settling rate.

Thus, the time variability of these systems, both in terms of the
pulsations of the white dwarf and in terms of the variable accretion
rate, allows us to probe the physics of accretion and gravitational
settling in these objects.

\acknowledgements M.H.M. and T.v.H are grateful for the financial
support of the National Science Foundation, under awards AST-0507639
and AST-0607480, respectively.  M.H.M. and S.E.T. gratefully
acknowledge the support of the Delaware Asteroseismic Research Center.


\clearpage

\end{document}